\documentclass[apj,12pt,preprint]{emulateapj}
\usepackage{graphicx}

\journalinfo{Accepted for publication in ApJ}
\shorttitle{Eclipsing dMe Binary LP~133-373}
\shortauthors{Vaccaro et al.}
\begin{document}
\title{LP~133-373: A New Chromospherically Active Eclipsing 
\lowercase{d}M\lowercase{e} Binary with a Distant, Cool White Dwarf  Companion}
 
\author{T.R. Vaccaro\altaffilmark{1}, M. Rudkin\altaffilmark{1}, A. Kawka\altaffilmark{2}, S. Vennes\altaffilmark{1},
T.D. Oswalt\altaffilmark{1}, I. Silver\altaffilmark{1}, M. Wood\altaffilmark{1}, and J. Allyn Smith\altaffilmark{3}}
\altaffiltext{1}{Department of Physics and Space Sciences and SARA Observatory, 150 W. University Blvd, Florida Institute of Technology, Melbourne, FL 32901, USA; tvaccaro, mrudkin, svennes, toswalt, isilver, wood@fit.edu.}
\altaffiltext{2}{Astronomick\'y \'ustav AV \v{C}R, Fri\v{c}ova 298,
 CZ-251 65 Ond\v{r}ejov, Czech Republic; kawka@sunstel.asu.cas.cz.}
\altaffiltext{3}{Department of Physics and Astronomy, Austin Peay State University, Clarksville, TN 37044  USA; smithj@apsu.edu.}

\begin{abstract}
We report the discovery of the partially eclipsing binary LP~133-373. Nearly 
identical eclipses along with observed photometric colors and spectroscopy 
indicate that it is a pair of chromospherically active dM4 stars in a circular 
1.6 d orbit. Light and velocity curve modeling to our differential photometry 
and velocity data show that each star has a mass and radius of 
$0.340\pm0.014 M_\odot$ and $0.33\pm0.02 R_\odot$. The binary is itself part 
of a common proper motion pair with LP~133-374 a cool DC or possible DA white 
dwarf with a mass of $0.49- 0.82\ M_\odot$, which would make the system at 
least 3 Gyr old.
\end{abstract}


\section{Introduction}
The star LP~133-373 (NLTT ~36188)---R.A.(2000)$=14^{\rm h}4^{\rm m}9^{\rm s}.0$, 
Dec.(2000)$=+50\arcdeg20\arcmin38\arcsec$---was originally listed in the 
\citet{luyt1979} catalog as a red star ($R=15.2$ mag) with a fainter 
($R=17.7$ mag) white dwarf star (LP~133-374, NLTT~36191) as a 
common-proper-motion companion separated by $5\arcsec$. Figure~\ref{fig1} 
shows a finder chart for these stars.

LP133-373 was found to be a partially eclipsing binary during unfiltered 
CCD-based time series photometry of the field in an attempt to detect 
variability of the white dwarf \citep{rud2003,osw2005}.  
Their original estimate of the binary period ($0.81$ d) corresponds to half 
the correct period ($1.63$ d) because the similarity of the stellar types 
yields nearly identical light curves for primary and secondary eclipses. This 
conclusion is supported by detailed light curve modeling and the splitting of 
emission and well as absorption lines near quadrature. 
Light curve variations outside of eclipse indicate dark surface spots, which 
is consistent with the chromospheric activity observed in the spectra (Balmer 
and \ion{Ca}{2} emissions).

Only a few eclipsing low-mass binaries in detached systems have been studied 
in detail.  \citet{lope2006a} and \citet{lope2006b} report 9 such systems. 
Other recent identifications were made by \citet{bay2006}, \citet{heb2006}, 
and \citet{you2006}. Late-type M dwarf stars are prominent members of the 
class of eclipsing cataclysmic variables. However, the late-type star is often 
outshined by the white dwarf's accretion disk providing little or no mass, 
luminosity, or radius information crucial in understanding the physical and 
evolutionary nature of these stars. Reviews of late-type mass-radius relations 
are given by \citet{cail1990}, \citet{chab2000}, and \citet{reid2000}.

We present in \S 2 new observations of the common proper-motion pair 
LP~133-373/374, and, in particular, photometric and spectroscopic data 
confirming that LP~133-373 is itself an eclipsing late-type binary. Detailed 
orbital and stellar properties of the binary LP~133-373 are derived in \S3.1 
and an analysis of the common proper-motion companion LP~133-374 is presented 
in \S3.2. We summarize and conclude in \S 4.

\section{Observations}

\subsection{Photometry}

We recently reprocessed the original photometric measurements obtained by 
\citet{smit1997} on 1994 April 2 at the Kitt Peak National Observatory (KPNO) 
0.9 m telescope. See  \citet{smit1997} for more details. We re-measured the 
BVRI colors for the red dwarf and white dwarf components, respectively: 
$B=16.907\pm0.029$ mag, $V=15.319\pm0.014$ mag, $R=14.093\pm0.012$ mag, 
$I=12.476\pm0.013$ mag, and $B=18.587\pm0.029$ mag, $V=18.020\pm0.053$ mag, 
$R=17.387\pm0.027$ mag, $I=16.274\pm0.029$ mag. The white dwarf colors were 
differentially determined with a field star and the red star's photometry was 
obtained with an aperture that excluded the white dwarf. The red dwarf data 
were obtained out of eclipse. The colors clearly show that the red star is 
approximately of dM4-5 spectral type in agreement with the spectra, which 
exhibit chromospheric activity and provide measurements of the TiO5 bandheads 
indicating a dM4 classification (see Fig.~\ref{fig2}) in agreement with 
previous classifications \citep{silv2005}. Our comparisons of the available 
colors with other dM stars in the literature \citep{cox2000,reid2000} also 
place it in the dM4 range, which we assume for our analysis.

\begin{figure}
\plotone{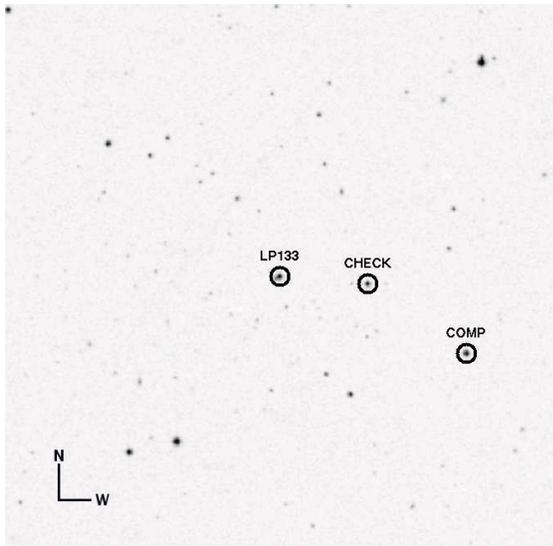}
\caption{Finder chart (red POSSII) for LP133-373/374 from The STScI Digitized 
Sky Survey indicating by circles the binary LP133-373, the comparison, and 
check stars. The distant white dwarf (LP133-374) is barely visible to the 
southeast of the brighter binary.  The image is $10\arcmin\times10\arcmin$.
\label{fig1}}
\end{figure}

\begin{figure*}
\plotone{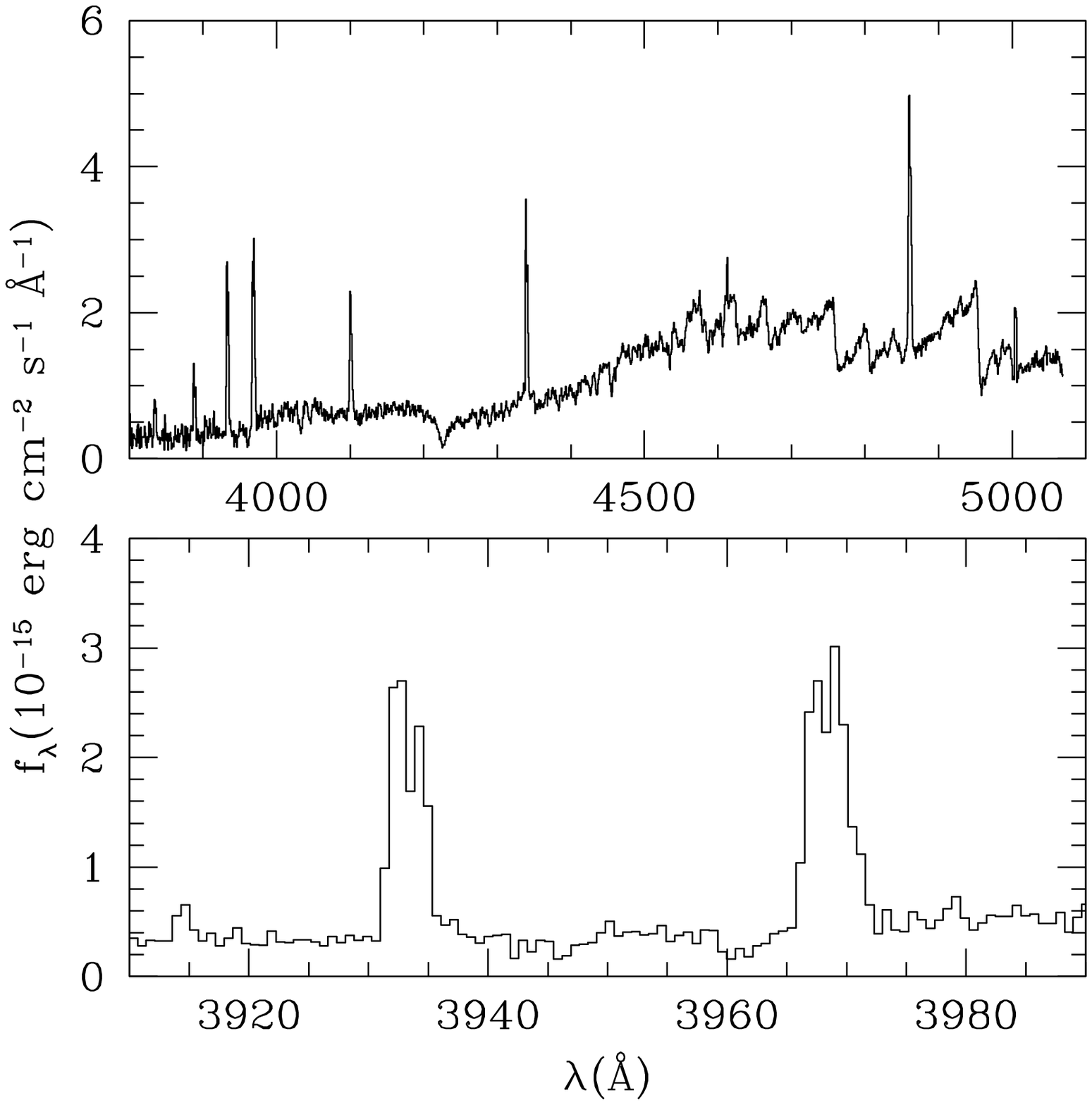}
\caption{({\it Top}) Mayall 4m KPNO spectrum of LP133-373 
($HJD = 2453580.703$), and ({\it bottom}) \ion{Ca}{2} H and K emission lines 
can be seen split as the binary was near quadrature.
\label{fig2}}
\end{figure*}

Series of images of LP133-373 were taken with the 0.9 m telescope of the 
Southeastern Association for Research in Astronomy (SARA), which is also 
located at KPNO. The initial eclipsing data were obtained in 2003 May using 
SARA's Ap7P CCD and no filters. Differential light curves were produced from 
the 2003 through 2005 observing seasons using various comparison stars in the 
field. Since the data were collected and reduced by several different 
observers using different comparison stars during this time and due to the 
unfiltered nature of the photometry, these light curves served only to 
establish the ephemeris. Observations made on the (UT) nights of 2006 May 2, 
3, 4, 21 and 30 were taken with SARA using the Finger Lakes CCD with Johnson 
R and I filters. A consistent comparison was used along with a check star 
that was always in the field to get differential measurements. Typical 
exposures were 30 seconds that yielded a ${\rm S}/{\rm N}\approx$40. These 
stars are shown in Figure~\ref{fig1}.

All new photometric data were reduced using standard procedures within 
IRAF\footnote{IRAF is distributed by the National Optical Astronomy 
Observatories, which are operated by the Association of Universities for 
Research in Astronomy, Inc., under cooperative agreement with the National 
Science Foundation.} and measured  with apertures a few arc seconds wide so 
as to exclude the white dwarf, which may have contributed less than one 
percent in R or I in some cases.  However, we believe that observations in 
these pass bands for the white dwarf reported by \citet{smit1997} are 
contaminated by LP133-373; therefore, we rely on the Sloan Digital Sky Survey 
(SDSS) ugriz photometry for the white dwarf, which are $u=20.018\pm0.038$ mag, 
$g=18.797\pm0.008$ mag, $r=18.256\pm0.007$ mag, $i=18.004\pm0.009$ mag, 
$z=18.076\pm0.020$ mag. Sloan data for the red dwarf were marked as saturated 
and could not be used.

Finally we obtained Two Micron All Sky Survey \citep[][2MASS]{skru2006} JHK photometry of the binary LP133-373: $J=10.905\pm0.020$, $H=10.306\pm0.021$, and $10.091\pm0.015$. The data are useful as an independent verification of the average M dwarf absolute luminosity. The start and end of the observations are HJD 2451322.699946 and HJD 2451322.704645, respectively, and according to our ephemeris (see \S 3.1 and Table~\ref{tbl1}) the data were obtained out of eclipse (phase$=0.7$). 

\begin{deluxetable}{cc}
\tablecaption{LP133-373 parameters ($q \equiv 1.0$)\label{tbl1}} 
\tablewidth{0pt}
\tablehead{
\colhead{Parameter} &   \colhead{Value}   \\ }
\startdata
$T_{0}$  & HJD $2452760.70502\pm0.00013$   \\ 
$P$ & $1.6279866\pm 0.0000004$  d \\ 
$a$ & $5.10 \pm 0.22\ R_\odot$   \\ 
$i$ & $85.3  \pm 0.05^\circ$  \\ 
$T_1$ & $3058\pm195$ K\\
$T_2$ & $3144\pm206$ K\\
$M_{\rm bol,1}$ & $10.0\pm0.5$ \\
$M_{\rm bol,2}$ & $9.8\pm0.5$ \\
$\Omega_{1}$ & $16.51\pm0.12$   \\
$\Omega_{2}$ & $16.34\pm 0.094$  \\
$R_1, R_2$ & $0.330\pm0.014\ R_\odot $ \\
$M_1, M_2$ & $0.34\pm0.02\ M_\odot$ \\
\enddata
\end{deluxetable}{}

\subsection{Spectroscopy}

\subsubsection{KPNO 1989 February}

Spectra of the white dwarf LP133-374  and red dwarf LP133-373 were originally obtained with the  Ritchey-Chr\'etien (RC) spectrograph attached to  the 4 m telescope at KPNO on 1989 February 7 (UT). We obtained a single exposure of 1200 s with both stars on the slit. The BL250 grating (158 lines mm$^{-1}$) and TI-2 CCD (15 $\micron$ pixel size, circa 1989) were used to obtain a spectral range of 3500 to 6200 \AA\ with a dispersion of 3.45 \AA\  pixel$^{-1}$ and a resolution of $\approx 14$ \AA. The resulting signal-to-noise ratio reached 15 in the red dwarf spectrum near 4150 \AA, and 5 throughout the white dwarf spectrum. The white dwarf spectrum appears featureless to the noise limit.  The spectrum is also relatively red indicating a  low effective temperature at which hydrogen Balmer lines are expected to be weak. The white dwarf is tentatively classified as a DC. A DA classification remains possible and would be confirmed with the acquisition of a high signal-to-noise ratio H$\alpha$ spectrum. The red dwarf spectrum revealed emission lines characteristic of chromospherically active stars. 

\subsubsection{KPNO 2005 July}

We obtained two optical spectra of LP133-373 with exposure times of 1800 s  on 2005 July 29 UT at the Mayall 4m telescope at KPNO.  We used the RC spectrograph using the BL450 grating in the second order resulting in a dispersion of 0.70 \AA\ pixel$^{-1}$ and a resolution of 1.8 \AA.  The 2k$\times$2k T2KB CCD camera with 24$\micron$ pixel size imaged the spectra.  An 8-mm CuSO$_4$ order-blocking filter was used to decrease the likelihood of order overlap within the blue end of the spectrum.  The range of wavelengths covered was 3800\AA\ to 5100\AA. We caught the stars near quadrature ($HJD = 2453580.67333196$ or phase$=0.67$,  and $HJD = 2453580.70308391$ or phase$=0.69$) as predicted by the ephemeris generated from the early photometric data. Figure~\ref{fig2} shows  prominent emission lines of \ion{H}{1} and \ion{Ca}{2} H and K. The  \ion{Ca}{2} H and K and the Balmer lines were split  indicating a velocity separation between the two binary components of $\approx 144$ km s$^{-1}$. Further details are presented in the analysis section (\S 3.1).

\subsubsection{APO 2006 September}

LP~133-373 was observed using the Dual Imaging Spectrograph (DIS) attached to the 3.5 m telescope at the Apache Point Observatory (APO) on 2006 September 30  02:31:24.7 UT and 02:47:16.0 UT (mid-exposure times). The exposure time of each spectrum is 900 s. We used the 830.8 line mm$^{-1}$ grating to obtain a spectral range of 6440 to 8150 \AA\ with a dispersion of 0.84 \AA\ pixel$^{-1}$ in the red. We also used the 1200 line mm$^{-1}$ grating to obtain a spectral range of 3830 to 5030 \AA\ with a  dispersion of 0.62  \AA\ pixel$^{-1}$ to obtain a spectrum in the blue. The slit was set for 1.5" resulting in a resolution of 2.1 \AA\ in  the red and 1.7 \AA\ in the blue. For each spectrum, the signal-to-noise ratio reached 5 near 4150 \AA\ and 20 near H$\alpha$.

Previous spectra had been obtained  at APO \citep{silv2002, silv2005} on 2001 February 4  09:59:13 UT (mid-exposure) but did not show double lines because  the orbit was near conjunction (phase$=0.9$).  Our more recent APO spectra (2006 September 30) taken nearly an hour after eclipse (phase$=0.53$)   yields an H$\alpha$ emission velocity $v({\rm H}\alpha)=-32.0\pm5.0$ km s$^{-1}$ which should closely match the systemic velocity but  the expected  line split ($\approx 20$ km s$^{-1}$) at this time was not detected due to limited resolution ($\approx 100$ km s$^{-1}$). Further details are presented in the analysis section (\S 3.1).

All new spectroscopic data were reduced using standard procedures within IRAF. Heliocentric corrections were applied to the wavelength scale of all spectra.

\section{Analysis}

\subsection{The eclipsing binary LP~133-373}

\begin{figure}
\plotone{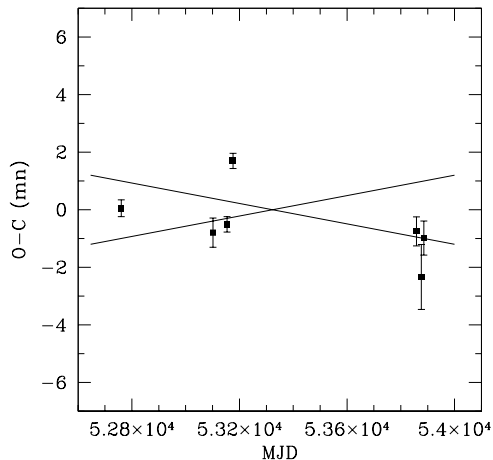}
\caption{$O-C$ diagram showing deviations of the epoch measurements (in units of minutes) of seven usable eclipses relative to the adopted ephemeris versus the epoch (MJD). The full lines show expected deviations from the ephemeris if the period is increased or decreased by 5$\sigma$ (0.000002 d). }\label{fig3}
\end{figure} 

To determine the binary parameters we followed four steps. First, we used the epoch of all seven eclipses, which include four eclipses  observed using unfiltered CCD data taken during 2003-2005 and three eclipses observed using filtered photometry in 2006 May, and we determined the initial epoch of the primary mid-eclipse and the orbital period (Table~\ref{tbl1}). Figure~\ref{fig3} shows the $O-C$ diagram for the epoch of all seven eclipses. Significant deviations with respect to our ephemeris are evident. We attribute these deviations to the presence of surface spots with varying contrasts and locations. The epoch of primary eclipse is (HJD)
\begin{displaymath}
T = 2452760.70502\pm0.00013 + E\  1.6279866\pm 0.0000004\  {\rm d}
\end{displaymath}

In a second step, we  used the ephemeris as a starting point for the binary star program PHOEBE \citep{prsa2005} and we performed detailed light curve modeling of our R and I data taken in 2006 May. We do not model unfiltered data. The program PHOEBE is based on the Wilson-Devinney (WD) program \citep{wils1971,wils1979,wils1990}. The similarity in the primary and secondary eclipses led us to adopt a mass ratio of 1.0, a circular orbit, and typical dM4 temperatures \citep{cox2000, reid2000} for each star of 3100 K.   

The radial velocity measurements provide the only anchor in establishing the semi-major axis $a$.  Radial velocities were determined from our 2005 July spectroscopy, which included measurements from H$\beta$, H$\gamma$, H$\delta$ and \ion{Ca}{2} K emission lines while excluding the very blended lines of H$\epsilon$ and \ion{Ca}{2} H.  The line centers were measured using IRAF's routine for de-blending multiple profiles, in this case two profiles---one for each star.  Only 2 sets of heliocentric velocities were obtained at phase$=0.67$, $v_1 = -108.2\pm4.4$ km~s$^{-1}$ and $v_2 = 31.1\pm4.1$ km~s$^{-1}$, and phase$=0.69$, $v_1 = -100.9\pm6.3$ km~s$^{-1}$ and $v_2 = 47.6\pm7.2$ km~s$^{-1}$.    The average velocity separation is $144\pm8$ km~s$^{-1}$. A best fit to the velocities starting with the known period, a systemic velocity between the measured velocities $\approx$ -33 km~s$^{-1}$ (which is close to the APO measurement of $v=-32.0\pm5.0$ km s$^{-1}$ at phase 0.53), and an inclination of 90$^\circ$ yields a minimum total mass of the system $\ga$0.67$M_\odot$. 

We repeated the radial velocity analysis using the absorption line spectra.  We adopted the spectrum of the dM3.5 star Gliese 15B as a template for both component stars of LP~133-373.  The spectrum of Gliese~15B was obtained at KPNO on 2006 November 27 with the same instrumental set-up used on 2005 July.  We established the zero-velocity scale using the radial velocity measurement of  \citet{nide2002}, $v({\rm Gl15B})=11.0\pm0.4$ km s$^{-1}$. We shifted the two templates independently within a velocity range of $-200$ to $+200$ km s$^{-1}$, and fitted the combined templates to the observed spectrum of LP~133-373 using a $\chi^2$ minimization technique. The best fits for the 2005 July spectra resulted in $v_1 = -94\pm10$ km s$^{-1}$ and $v_2 = 46\pm10$ km s$^{-1}$ (phase$=0.67$), and $v_1 = -90\pm10$ km s$^{-1}$ and $v_2 = 60\pm10$ km~s$^{-1}$ (phase$=0.69$), which corresponds to an average velocity separation of $145\pm14$ km~s$^{-1}$. The velocity separation measurements using absorption and emission line spectra are essentially identical. We also measured the systemic velocity in the APO 2006 September spectrum ($-24\pm 7$ km~s$^{-1}$) using \ion{K}{1} resonance lines at $\lambda=7664.911$ and $7698.974$ \AA. We find that the mass ratio ($M_2/M_1$) ranges from $\approx 1.0$ using the emission spectra to $\approx 1.1$ using the absorption spectra. Although we adopted $M_2/M_1=1$ we will explore the effect of a varying mass ratio (see below).

Next, in a third step, the inclination and potentials (stellar radii) were then iteratively adjusted within PHOEBE until both eclipse depths and widths matched the observed light curve. Note that all solutions are based on a atmosphere of 3500 K ramped from a black body approximation at 1500 K, and on logarithmic limb darkening with coefficients by \citet{vanh1993} for temperatures of 3500 K since the coefficient values are not known below this temperature.  Table~\ref{tbl1} presents the simultaneous light/velocity solutions. 

\begin{figure}
\plotone{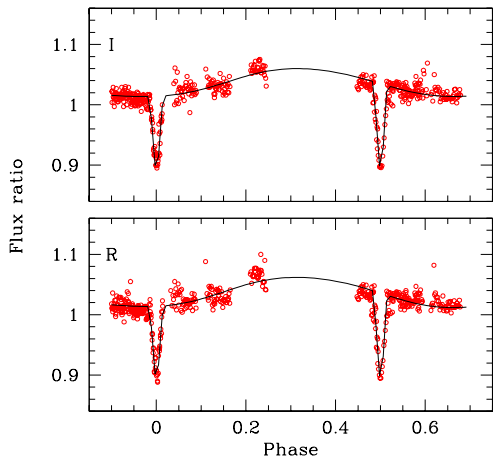}
\caption{Light curve model ({\it solid lines}) compared with our I-band (top) and R-band ({\it bottom}) photometry ({\it open circles}).
\label{fig4}}
\end{figure}

\begin{figure*}
\plotone{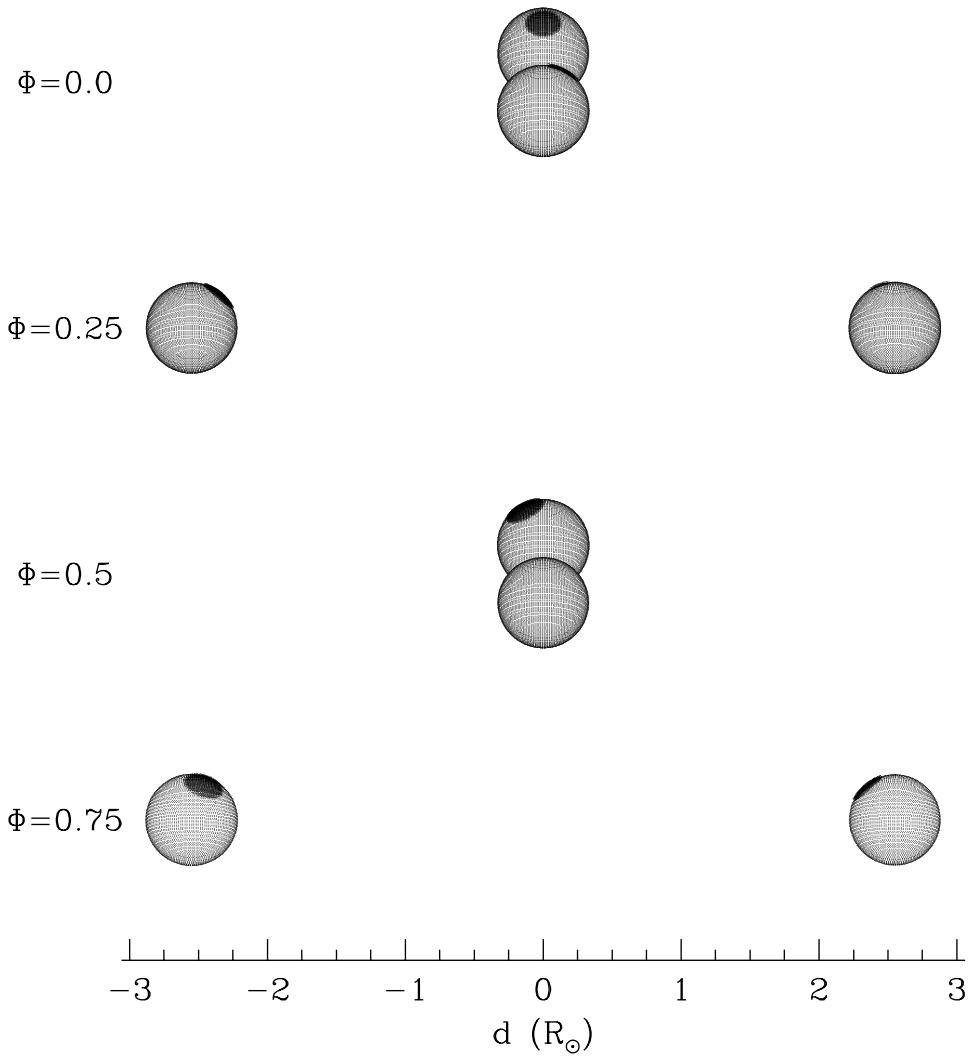}
\caption{Binary configuration as seen along the line of sight at $\Phi = 0.0, 0.25$, and $0.5$. The images were generated with PHOEBE.
\label{fig5}}
\end{figure*}

Finally, in a last fourth step we introduced surface spots.  Initial fits were done with the light levels adjusted to match the eclipses. We subsequently scaled the model light curve to values outside of eclipse (near phase 0.25) and spots were added to fit the complete light curve since a spot wave is clearly evident.  Light curve models were compared to the eclipse geometry implied by our $R$, $I$, and unfiltered photometry.  The unfiltered comparisons are only for a check on the ephemeris since spot configurations are surely different between the 2003 and 2006 epochs.  We fit our photometric data with model light curves using  the spot parameters given in Table~\ref{tbl2}. No satisfactory fit was possible without starspots, so cool spots were added one at a time, with typical temperature factors, $\frac{T_{spot}}{T_{photosphere}}$, of about 0.80. Spots on the facing hemisphere of each star were needed. A similar effect on the light curves would occur if the spots were on the outer facing hemispheres but the models were best fit with the inner facing configuration. The spot parameters were refined with the differential corrections routine part of the Wilson-Devinney code. Figure~\ref{fig4} shows the light curve using   parameters for the two spots   given  in Table~\ref{tbl2}, and Figure~\ref{fig5} shows the binary configuration at four phases ($\Phi =0.0, 0.25, 0.5, 0.75$)  and as seen along the line of sight.

\begin{deluxetable}{ccccc}
\tablecaption{LP133-373 spot solutions in 2006 May data\label{tbl2}} 
\tablewidth{0pt}
\tablehead{
\colhead{Spot} & \colhead{Co-Latitude} & \colhead{Longitude} &  \colhead{Ang. Radius}  & \colhead{Temp. Factor}   \\
 \colhead{\#} &   \colhead{(radians)}       & \colhead{(radians)} &  \colhead{(radians)}   & \colhead{$\frac{T_{spot}}{T_{photosphere}}$}      \\
}
\startdata 
1 & 0.70$\pm$0.17 & 0.00$\pm$0.04 & 0.40$\pm$0.02 & 0.72$\pm$0.06\\
2 & 0.55$\pm$0.07 & 5.30$\pm$0.06 & 0.45$\pm$0.03 & 0.78$\pm$0.05\\  
\enddata
\end{deluxetable}{}

We also examined the sensitivity of the solutions to our our assumption of a mass ratio $q=1$. If we let the mass ratio reach $q=0.9$ and 1.1, in both cases the separation increases from 5.1 to 5.15 $R_\odot$ corresponding to a slight increase in the total systemic mass from 0.68 to 0.69 $M_\odot$ and no significant effect on the inclination. Of course, by introducing a mass asymmetry, the predicted systemic velocity shifts from  $-33$  to  $-29$ km~s$^{-1}$ at $q=0.9$ and to $-40$ km~s$^{-1}$ at $q=1.1$. The solutions closely match the limits allowed by our systemic velocity measurement of $v=-32.0\pm5.0$ km s$^{-1}$, and, therefore, the actual mass ratio should be found within the range $q=0.9-1.1$.

Tables~\ref{tbl3} and \ref{tbl4} present our photometry, where $\Delta R$, $\Delta I$ refer to magnitude differences between LP133-373 and the comparison star shown in Figure~\ref{fig1}.

\subsubsection{Luminosity and distance estimate of the red dwarf}

We computed the distance to the binary several ways. This allowed us to check the modeled binary luminosity for consistency and to find the luminosity of the white dwarf common proper motion component. Two methods of determining luminosity for the binary from observable measurements were used.  One method uses the V-I colors to obtain the absolute visual magnitude $M_V$. The other method utilizes the TiO5 band strength to get $M_V$.  The apparent magnitude ($m_V$) can then be used to compute a distance modulus.  We assumed that the stars were identical enough to facilitate a simple distance correction.  The inverse square law allows us to correct the distance computed for two identical stars by treating them as one and multiplying the result by $\sqrt{2}$.

The empirical color magnitude diagram of nearby stars yields the following relationship for $0.85 < V-I < 2.85$ \citep{reid2000}
\begin{displaymath}
M_V=3.98+1.437\,(V-I)+1.073\,(V-I)^2-0.192\,(V-I)^3,
\end{displaymath}
with $\sigma(M_V) =  0.5$.  Using our (V-I)=2.843 we get an $M_V$=12.3, which gives a distance of 40 pc for a single star of this color and observed $V=15.319$. However, we correct this distance by a factor of $\sqrt{2}$ and get the  distance to the system as 56 pc.

A similar relationship exists for the TiO5 band strength \citep{reid1995},
\begin{displaymath}
M_V=25.33-53.15\,({\rm TiO5})+64.1\,({\rm TiO5})^2-29.14\,({\rm TiO5})^3,
\end{displaymath}
with $\sigma(M_V) = 0.5$.  Using TiO5$=0.37$ \citep{silv2002} we get $M_V=13.0$ and a distance of 30 pc, which when corrected for two stars becomes 42 pc.

The PHOEBE light curve models compute bolometric magnitudes for both stars, which are similar ($M_{\rm bol} \approx 9.9\pm0.5$).  Bolometric corrections for the visual bandpass can be computed and subtracted from the modeled $M_{\rm bol}$ to give $M_V$ using the following relation \citep{reid2000}
\begin{displaymath}
BC_V=0.27-0.604\,(V-I)-0.125\,(V-I)^2,
\end{displaymath}
which gives us  $BC_V=2.5$ and therefore  $M_V=12.4$ for a corrected distance of 55 pc.  The model is consistent with the two independently determined distance estimates ranging from  42 to 56 pc.

In summary, the distance estimates imply a modulus of
\begin{displaymath}
m-M = 3.4\pm0.3
\end{displaymath}
Adopting $m_K-M_K = 3.4\pm0.3$,  the absolute magnitude $M_K = 6.7\pm0.3$ for the pair or $M_K = 7.4\pm0.3$ for each binary component assuming equal luminosity. The measured $V-K= 5.23\pm0.03$ and our $M_K$ estimate for LP~133-373 are consistent  with established relations for nearby stars  \citep{reid2000}. The main limitation to the usefulness of such relations is the intrinsic scatter found  in color-luminosity  measurements. Note that the absolute JHK magnitudes of the nearby white dwarf companion (see \S 3.2) are $\ga 13$ mag; Therefore, potential contamination of the infrared data  by the white dwarf is insignificant.

\subsection{Physical parameters and age of the white dwarf LP~133-374}

\begin{figure}
\plotone{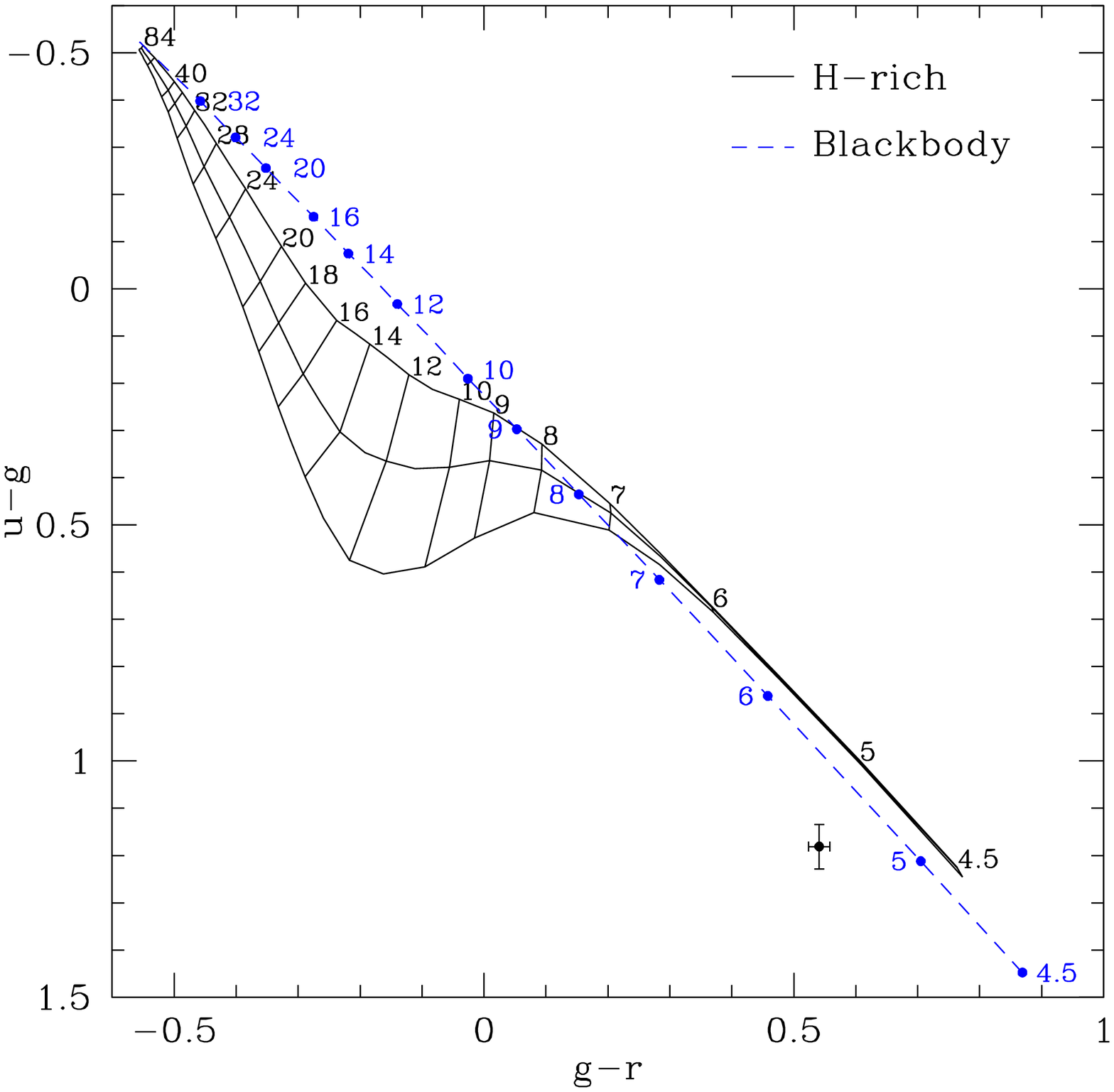}
\caption{SDSS $(u-g)$ vs. $(g-r)$ photometry of LP133-374 ({\it full circle} with error bars) compared to synthetic colors of hydrogen-rich white dwarfs ({\it full line}) and blackbody ({\it dotted line}). The effective temperature is indicated in units of 1000 K and for hydrogen-rich colors $\log{g} = 7.0$, 8.0 and 9.0 ({\it bottom to top}). The grid shown does not include the effect of missing blue/ultraviolet opacity.
\label{fig6}}
\end{figure}

\begin{figure}
\plotone{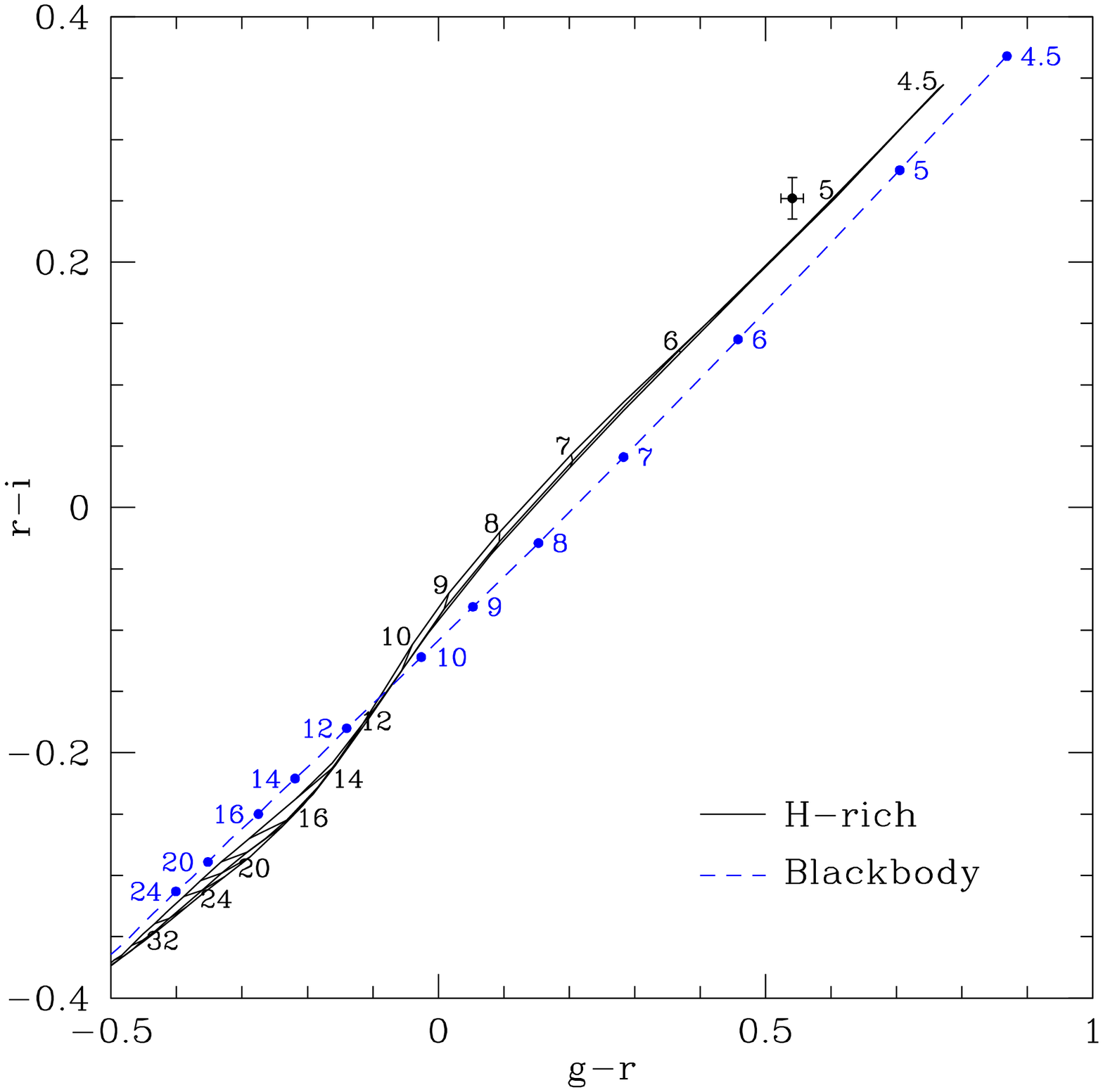}
\caption{SDSS $(r-i)$ vs. $(g-r)$ photometry of LP133-374 compared ({\it full circle} and error bars) to synthetic colors of hydrogen-rich white dwarfs ({\it full line}) and blackbody ({\it dotted line}). The effective temperature is indicated in units of 1000 K and for hydrogen-rich colors $\log{g} = 7.0$, 8.0 and 9.0 ({\it bottom to top}).
\label{fig7}}
\end{figure}

To analyze the photometric and spectroscopic data  of the white dwarf LP133-374, we calculated a set of synthetic SDSS $ugriz$ colors using a grid of pure-hydrogen models \citep{kaw2006} for $T_{\rm eff} = 4500$ to 84\,000 K and $\log{g} = 7.0$, 8.0 and 9.0. We also calculated synthetic $ugriz$ colors for black-body spectra with temperatures ranging from  $T_{\rm eff} = 4500$ to 84\,000 K. Figures~\ref{fig6} and \ref{fig7} show the observed $(u-g)$ versus $(g-r)$ and $(r-i)$ versus $(g-r)$ colors, respectively, compared to the synthetic colors for pure-hydrogen models and the black-body colors. 

Note that Kowalski \& Saumon (2006) recently established that the extended line wing of Ly$\alpha$ contributes significantly to the total opacity in the blue part of the optical spectrum. The additional opacity is due to perturbation of the hydrogen atom by neighboring H atoms and H$_2$ molecules. An examination of Figure 2 in Kowalski \& Saumon (2006) suggests that the effect of this additional opacity on a 5800 K hydrogen-rich white dwarf  corresponds to an increase of $\approx0.25$ and 0.02 mag in the $u$ and $g$ bands, respectively. Hence, the effect is mostly apparent in the $u-g$ color index and correspond, as observed, to a downward vertical shift in the $u-g$ versus $g-r$ diagram. 

The temperatures were derived by minimizing the $\chi^2$ between the observed photometry and the synthetic colors. The synthetic $u$ and $g$ magnitudes for hydrogen-rich models were corrected by $+0.25$ and $+0.02$ mag, respectively. The errors in temperatures were determined from considering the uncertainties in the observed colors only. Therefore, the quoted errors do not take into account systematic errors in  synthetic colors such as discussed above. The observed $(u-g)$ versus $(g-r)$ colors correspond to an effective temperature of $5200\pm100$ K using the pure-hydrogen sequence, and $5300\pm100$ K using the black-body colors. The observed $(r-i)$ versus $(g-r)$ colors correspond to an effective temperature of $5100\pm100$ K using the pure-hydrogen sequence and $5400\pm50$ K using black-body colors. 

\begin{figure*}
\centering
\includegraphics[viewport=18 15 585 300,clip,width=1.0\textwidth]{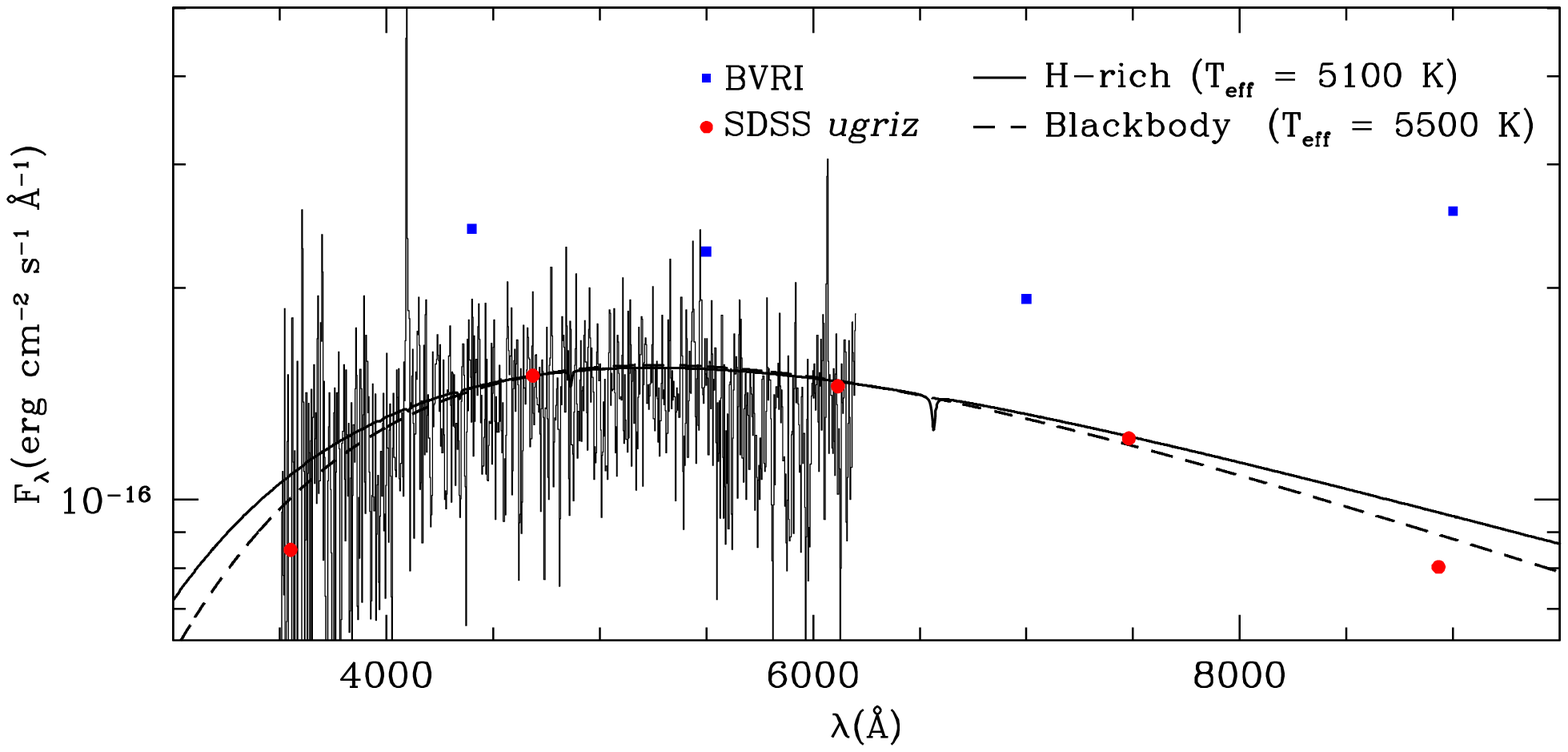}
\caption{SDSS {\it ugriz} and {\it BVRI} photometry and a spectrum of LP133-374  compared to a model DA spectrum and a black-body spectrum. The hydrogen model shown does not include the effect of missing blue/ultraviolet opacity.
\label{fig8}}
\end{figure*}

We also fit the SDSS {\it ugriz} photometry to synthetic {\it ugriz} absolute magnitudes, and found that $T_{\rm eff} = 5300\pm 200$ K when using the hydrogen-rich sequence (assuming $\log{g} = 8.0$) and $T_{\rm eff} = 5500\pm200$ K assuming a black-body. We also fit the available spectrum (3800 to 6190 \AA) to DA spectra at $\log{g} = 8.0$ and black-body spectra to obtain $T_{\rm eff} = 5100\pm200$ K and $T_{\rm eff} = 5580\pm160$ K, respectively. Figure~\ref{fig8} shows the spectrum and {\it ugriz} photometry of LP133-374 compared to a hydrogen-rich spectrum at $T_{\rm eff} = 5100$ K and a blackbody spectrum at $T_{\rm eff} = 5500$ K.

We determined the possible range of mass values for the white dwarf assuming both hydrogen-rich atmospheres and helium-rich atmospheres (using a black-body approximation). First, we calculated the absolute magnitude of the white dwarf based on the distance, and therefore if the system is between 42 and 56 pc, then the absolute magnitude ($M_g$) of the white dwarf is between 15.68 and 15.06. 

Assuming the white dwarf is hydrogen-rich, then the mean of the temperatures determined above is $T_{\rm eff} = 5175\pm100$ K. The mass range can then be determined using the mass-radius relations of  \citet{ben1999} with a hydrogen envelope of $M_H/M_\star = 10^{-4}$ and a metallicity of $Z=0$, where for a given temperature, the absolute magnitude is a function of the radius and therefore the mass. We determined a   mass   of $0.49 - 0.70\ M_\odot$ with a cooling age ranging from $3.6\times 10^9$ to $4.5\times 10^9$ years.

Assuming the white dwarf is helium-rich, and using a black-body approximation for the energy distribution, then the mean of the temperatures determined using the different methods as discussed above is $T_{\rm eff} = 5445\pm120$ K. For the mass determinations in this case, we used the mass-radius relations
of \citet{ben1999} without a hydrogen envelope and with a metallicity of $Z=0.001$. Therefore, if LP133-374 is a helium-rich white dwarf with $T_{\rm eff} = 5445$ K, we can estimate the mass to be between 0.55 and $0.82 M_\odot$ and the cooling age to be between $2.7\times 10^9$ and $4.5\times 10^9$ years.  The progenitor of a $0.82\, M_\odot$ white dwarf is a early-type star of $\approx 4\ M_\odot$ \citep{weid2000} with a main-sequence life-time $\approx 10^8$ years. These ages imply a minimum age of $3\times 10^9$ years for the LP133-373/374 system.

For a temperature range of 5000-5500 K, the corresponding 2MASS absolute magnitude $M_J = 13.4-13.6$, or an apparent magnitude of $m_J = 16.5-16.7$, fainter than the limiting magnitude of the 2MASS survey.

\section{Summary and conclusions}

We have modeled light and velocity curves of LP133-373 as an eclipsing binary with two similar dM type stars with spots.  Although the masses and radii of $0.340\pm0.014 M_\odot$ and $0.33\pm0.02 R_\odot$ make them appear marginally more massive and smaller than other stars of the same type  tabulated by \citet{reid2000}, a complete radial velocity study and additional light curves are required before we can reach a conclusion on this matter.  The uncertainty in the mass ratio and the paucity of velocity data create a large uncertainty in the semi-major axis and ultimately in the individual masses.  The role of spots also play a critical role in light curve modeling and the resulting fits.  The 2006 May data set shows light curve variations suggesting that spot activity had significantly changed within a month.  We intend on obtaining   high-dispersion    spectra of LP133-374 sampling a complete orbital period in the H$\alpha$ region. New photometry, covering several complete orbital periods, would be helpful in constraining the geometry of the spots and the individual temperatures. The new data will provide better mass, luminosity, and radius information that is so important for understanding low mass stars.   In addition, limb darkening of these cool stars is also important to the light curve models. A study of additional  stars like these is needed to refine limb darkening models. 

The white dwarf is tentatively classified as a DC spectral type, and a possible DA type. It has a temperature of 5100 to 5500 K depending on the atmospheric abundance of hydrogen. Adopting the distance of its common proper-motion companion LP133-373 (42-56 pc) we constrain the radius, hence the mass of the white dwarf. The mass of the white dwarf is estimated to be $0.49 \la M/M_\odot \la 0.82$. Consequently, the minimum total age of the system is 3 Gyr. The angular separation of 5$\arcsec$ between the white dwarf and the eclipsing binary corresponds to a projected separation of  210-280 AU. This separation excludes the possibility of  past interactions. 

New H$\alpha$ spectroscopy will also help establish the spectral type of the white dwarf, and accurately determine its physical characteristics. In turn, this will be helpful in establishing the age of the triple system from the white dwarf cooling age.
 
\acknowledgements

We thank KPNO  for a generous allocation of observing time.  We are especially grateful for the observing support of Rob Wilkos, John Robertson, and Kyle Johnston.   M.R. is supported by a NASA Graduate Student Research Program NGT5-50450. A. Kawka is supported by GA \v{C}R 205/05/P186. S.V. acknowledges support from the College of Science at the Florida Institute of Technology. This work was supported in part by NSF grant AST-0206115 (T.\ D.\ Oswalt Principal Investigator) to Florida Institute of Technology. M.W. acknowledges support from a NSF grant AST-0097616. We thank S. Shaw, R. Wilson, and Casey T. Hendley for useful discussions and N. Silvestri for sharing observing time at APO. We thank the anonymous referee for several useful comments that improved the paper.

This research is also based in part on observations obtained with the Apache Point Observatory 3.5-meter telescope, which is owned and operated by the Astrophysical Research Consortium (ARC).
Funding for the SDSS and SDSS-II has been provided by the Alfred P. Sloan Foundation, the Participating Institutions, the National Science Foundation, the U.S. Department of Energy, the National Aeronautics and Space Administration, the Japanese Monbukagakusho, the Max Planck Society, and the Higher Education Funding Council for England. The SDSS Web Site is http://www.sdss.org/.
This publication also makes use of data products from the Two Micron All Sky Survey, which is a joint project of the University of Massachusetts and the Infrared Processing and Analysis Center/California Institute of Technology, funded by the National Aeronautics and Space Administration and the National Science Foundation.

\clearpage

\LongTables
\begin{deluxetable}{cccccccccc}
\tablecaption{Photometric $R$ Data for LP133-373\label{tbl3}} 
\tablewidth{0pt}
\tablehead{
\colhead{HJD} & \colhead{$\Delta R$} & \colhead{HJD} &  \colhead{$\Delta R$}  & \colhead{HJD} & \colhead{$\Delta R$} & \colhead{HJD} & \colhead{$\Delta R$} & \colhead{HJD} & \colhead{$\Delta R$} \\
 \colhead{-2453800} &          & \colhead{-2453800} &               & \colhead{-2453800} &            & \colhead{-2453800} &                & \colhead{-2453800} &      \\
}
\startdata                                             
  57.74370 &   -0.024 &   57.86909 &   -0.005 &   58.70187 &   -0.027 &   59.67035 &   -0.037 &   59.98506 &   -0.058 \\ 
  57.74743 &   -0.004 &   57.86968 &   -0.018 &   58.70397 &   -0.042 &   59.67243 &   -0.035 &   59.98715 &   -0.057 \\ 
  57.74804 &   -0.012 &   57.87225 &   -0.009 &   58.70605 &   -0.037 &   59.67451 &   -0.054 &   59.98923 &   -0.094 \\ 
  57.74863 &   -0.016 &   57.87284 &   -0.014 &   58.70816 &   -0.043 &   59.67660 &   -0.017 &   59.99132 &   -0.045 \\ 
  57.75122 &   -0.015 &   57.87343 &   -0.058 &   58.71025 &   -0.049 &   59.67869 &   -0.039 &   59.99341 &   -0.044 \\ 
  57.75182 &   -0.018 &   57.87600 &   -0.006 &   58.71235 &   -0.032 &   59.68079 &   -0.035 &   76.65658 &    0.001 \\ 
  57.75241 &   -0.013 &   57.87659 &   -0.010 &   58.71443 &   -0.040 &   59.68287 &   -0.043 &   76.65862 &   -0.008 \\ 
  57.75499 &   -0.024 &   57.87718 &   -0.016 &   58.71653 &   -0.049 &   59.68495 &   -0.008 &   76.66048 &    0.000 \\ 
  57.75559 &    0.003 &   57.87975 &   -0.005 &   58.71862 &   -0.053 &   59.68705 &   -0.045 &   76.66320 &    0.010 \\ 
  57.75618 &   -0.018 &   57.88034 &   -0.019 &   58.72072 &   -0.036 &   59.68913 &   -0.004 &   76.66591 &   -0.003 \\ 
  57.75877 &    0.005 &   57.88093 &   -0.016 &   58.72281 &   -0.040 &   59.69123 &   -0.018 &   76.66863 &    0.013 \\ 
  57.75938 &   -0.014 &   57.88354 &   -0.011 &   58.72491 &   -0.036 &   59.69331 &   -0.028 &   76.67134 &    0.025 \\ 
  57.75998 &   -0.020 &   57.88414 &   -0.008 &   58.72700 &   -0.034 &   59.69539 &   -0.020 &   76.67406 &    0.060 \\ 
  57.76255 &   -0.015 &   57.88473 &    0.005 &   58.72910 &   -0.052 &   59.69747 &   -0.025 &   76.67678 &    0.063 \\ 
  57.76314 &    0.005 &   57.88732 &    0.002 &   58.73119 &   -0.031 &   59.69956 &   -0.011 &   76.67951 &    0.092 \\ 
  57.76374 &   -0.010 &   57.88792 &   -0.009 &   58.73329 &   -0.031 &   59.70165 &   -0.026 &   76.68222 &    0.102 \\ 
  57.76630 &   -0.027 &   57.88853 &   -0.001 &   58.73537 &   -0.035 &   59.70375 &   -0.027 &   76.68494 &    0.114 \\ 
  57.76689 &   -0.005 &   57.89112 &   -0.012 &   58.73746 &   -0.040 &   59.70584 &   -0.013 &   76.68766 &    0.120 \\ 
  57.76749 &   -0.022 &   57.89171 &   -0.002 &   58.73956 &   -0.032 &   59.70793 &   -0.027 &   76.69039 &    0.120 \\ 
  57.77005 &   -0.017 &   57.89231 &   -0.011 &   58.74165 &   -0.030 &   59.71002 &   -0.023 &   76.69312 &    0.110 \\ 
  57.77064 &   -0.023 &   57.89490 &   -0.006 &   58.74375 &   -0.026 &   59.71212 &   -0.024 &   76.69584 &    0.098 \\ 
  57.77124 &   -0.024 &   57.89550 &   -0.001 &   58.74583 &   -0.028 &   59.71420 &   -0.018 &   76.69856 &    0.086 \\ 
  57.77383 &   -0.024 &   57.89611 &   -0.008 &   58.74793 &   -0.038 &   59.71629 &   -0.023 &   76.70128 &    0.057 \\ 
  57.77443 &   -0.022 &   57.89870 &    0.008 &   58.75002 &   -0.049 &   59.71837 &   -0.013 &   76.70400 &    0.049 \\ 
  57.77504 &   -0.014 &   57.89929 &    0.001 &   58.75212 &   -0.038 &   59.72045 &   -0.026 &   76.70672 &    0.029 \\ 
  57.77763 &   -0.025 &   57.89989 &   -0.016 &   58.75421 &   -0.039 &   59.72254 &   -0.023 &   76.70944 &    0.014 \\ 
  57.77822 &   -0.024 &   57.90248 &   -0.003 &   58.75632 &   -0.040 &   59.72463 &   -0.037 &   76.71216 &    0.002 \\ 
  57.77882 &   -0.023 &   57.90309 &   -0.013 &   58.75841 &   -0.051 &   59.72672 &   -0.008 &   76.71488 &    0.002 \\ 
  57.78141 &   -0.023 &   57.90369 &   -0.015 &   58.76051 &   -0.017 &   59.72880 &   -0.025 &   76.71760 &   -0.002 \\ 
  57.78200 &   -0.015 &   57.90628 &   -0.004 &   58.76260 &   -0.014 &   59.73090 &   -0.028 &   76.72033 &   -0.008 \\ 
  57.78260 &   -0.025 &   57.90688 &   -0.011 &   58.76468 &   -0.011 &   59.73299 &   -0.036 &   76.72665 &   -0.002 \\ 
  57.78520 &   -0.019 &   57.90747 &   -0.002 &   58.76678 &    0.012 &   59.73508 &   -0.028 &   76.73270 &   -0.004 \\ 
  57.78580 &   -0.036 &   57.91008 &   -0.006 &   58.76889 &    0.041 &   59.73718 &   -0.024 &   76.76273 &   -0.010 \\ 
  57.78639 &   -0.032 &   57.91068 &   -0.005 &   58.77098 &    0.044 &   59.73928 &   -0.012 &   76.76536 &   -0.023 \\ 
  57.78895 &   -0.027 &   57.91128 &   -0.017 &   58.77307 &    0.068 &   59.77400 &   -0.092 &   76.76809 &   -0.016 \\ 
  57.78954 &   -0.017 &   57.91386 &   -0.009 &   58.80441 &   -0.042 &   59.77940 &   -0.037 &   76.77079 &   -0.015 \\ 
  57.79014 &   -0.012 &   57.91446 &   -0.012 &   58.80650 &   -0.035 &   59.78218 &   -0.030 &   76.77348 &   -0.016 \\ 
  57.79273 &   -0.027 &   57.91506 &   -0.003 &   58.80859 &   -0.020 &   59.78426 &   -0.026 &   76.77619 &   -0.009 \\ 
  57.79332 &   -0.033 &   57.91766 &   -0.002 &   58.81069 &   -0.039 &   59.78636 &   -0.050 &   76.77891 &   -0.016 \\ 
  57.79392 &   -0.029 &   57.91826 &   -0.005 &   58.81278 &   -0.033 &   59.78844 &   -0.035 &   76.78164 &   -0.043 \\ 
  57.79653 &   -0.028 &   57.91885 &   -0.012 &   58.81488 &   -0.031 &   59.79054 &   -0.047 &   76.78436 &   -0.026 \\ 
  57.79713 &   -0.027 &   57.92145 &   -0.006 &   58.81697 &   -0.038 &   59.79263 &   -0.034 &   76.78707 &   -0.014 \\ 
  57.79772 &   -0.029 &   57.92204 &   -0.003 &   58.81908 &   -0.035 &   59.79472 &   -0.050 &   76.78979 &   -0.015 \\ 
  57.80032 &   -0.008 &   57.92264 &   -0.014 &   58.82118 &   -0.028 &   59.79680 &   -0.045 &   76.79250 &   -0.029 \\ 
  57.80091 &   -0.025 &   57.92643 &   -0.015 &   58.82327 &   -0.031 &   59.79889 &   -0.017 &   76.79523 &   -0.016 \\ 
  57.80152 &   -0.012 &   57.92903 &   -0.013 &   58.82536 &   -0.026 &   59.80099 &   -0.009 &   76.79794 &   -0.022 \\ 
  57.80411 &    0.007 &   57.92962 &   -0.013 &   58.82746 &   -0.024 &   59.80308 &   -0.020 &   76.80068 &   -0.024 \\ 
  57.80471 &   -0.019 &   57.93021 &    0.009 &   58.82954 &   -0.037 &   59.80516 &   -0.027 &   76.80340 &   -0.015 \\ 
  57.80531 &   -0.012 &   57.93278 &   -0.011 &   58.83164 &   -0.034 &   59.80723 &   -0.019 &   76.80612 &   -0.023 \\ 
  57.80789 &   -0.011 &   57.93339 &    0.008 &   58.83374 &   -0.033 &   59.80930 &   -0.025 &   76.80884 &   -0.020 \\ 
  57.80850 &   -0.038 &   57.93399 &    0.005 &   58.83581 &   -0.035 &   59.81138 &   -0.040 &   76.81156 &   -0.025 \\ 
  57.80910 &   -0.005 &   57.93658 &    0.005 &   58.83791 &   -0.038 &   59.81347 &   -0.045 &   76.81427 &   -0.024 \\ 
  57.81170 &   -0.012 &   57.93717 &    0.002 &   58.84001 &   -0.040 &   59.81556 &   -0.028 &   76.81699 &   -0.024 \\ 
  57.81230 &   -0.014 &   57.93777 &    0.005 &   58.84210 &   -0.030 &   59.81765 &   -0.044 &   76.81971 &   -0.025 \\ 
  57.81289 &   -0.018 &   57.94036 &    0.001 &   58.84421 &   -0.037 &   59.81975 &   -0.043 &   76.82243 &   -0.026 \\ 
  57.81549 &   -0.012 &   57.94095 &   -0.018 &   58.84630 &   -0.031 &   59.82183 &   -0.033 &   76.82515 &   -0.010 \\ 
  57.81609 &   -0.009 &   57.94156 &    0.000 &   58.84840 &   -0.039 &   59.82393 &   -0.039 &   76.82788 &   -0.013 \\ 
  57.81669 &   -0.026 &   57.94415 &   -0.027 &   58.85049 &   -0.026 &   59.82601 &   -0.047 &   76.83060 &   -0.006 \\ 
  57.81929 &    0.003 &   57.94475 &   -0.005 &   58.85258 &   -0.026 &   59.82810 &   -0.024 &   76.87240 &   -0.008 \\ 
  57.81988 &   -0.001 &   57.94534 &   -0.017 &   58.85469 &   -0.015 &   59.83020 &   -0.029 &   76.87515 &   -0.017 \\ 
  57.82049 &   -0.026 &   57.94793 &    0.005 &   58.85679 &   -0.031 &   59.83228 &   -0.040 &   76.87787 &   -0.020 \\ 
  57.82308 &   -0.033 &   57.94853 &    0.000 &   58.85888 &   -0.034 &   59.83437 &   -0.022 &   76.88057 &   -0.015 \\ 
  57.82367 &   -0.018 &   57.94914 &    0.015 &   58.86098 &   -0.034 &   59.83645 &   -0.013 &   76.88331 &   -0.086 \\ 
  57.82426 &   -0.012 &   57.95173 &    0.047 &   58.86307 &   -0.023 &   59.83853 &   -0.026 &   76.88603 &   -0.035 \\ 
  57.82685 &   -0.016 &   57.95233 &    0.034 &   58.86516 &   -0.037 &   59.84063 &   -0.027 &   76.88876 &   -0.038 \\ 
  57.82745 &   -0.018 &   57.95292 &    0.036 &   58.86725 &   -0.031 &   59.84271 &   -0.024 &   76.89148 &   -0.031 \\ 
  57.82804 &   -0.011 &   57.95551 &    0.071 &   58.86934 &   -0.020 &   59.84479 &   -0.026 &   76.89421 &   -0.025 \\ 
  57.83064 &   -0.008 &   57.95610 &    0.063 &   58.87145 &   -0.031 &   59.84689 &   -0.032 &   76.89693 &   -0.031 \\ 
  57.83124 &   -0.031 &   57.95671 &    0.061 &   58.87354 &   -0.039 &   59.84898 &   -0.022 &   76.89966 &   -0.024 \\ 
  57.83183 &   -0.013 &   57.95930 &    0.088 &   58.87565 &   -0.028 &   59.85107 &   -0.022 &   76.90238 &   -0.018 \\ 
  57.83442 &   -0.030 &   57.95990 &    0.107 &   58.87775 &   -0.032 &   59.85315 &   -0.028 &   76.90511 &   -0.009 \\ 
  57.83502 &   -0.011 &   57.96050 &    0.110 &   58.87984 &   -0.023 &   59.85523 &   -0.026 &   76.90783 &   -0.016 \\ 
  57.83561 &   -0.015 &   57.96308 &    0.080 &   58.88192 &   -0.025 &   59.85733 &   -0.029 &   76.91055 &   -0.014 \\ 
  57.83820 &   -0.031 &   57.96368 &    0.100 &   58.88403 &   -0.026 &   59.85941 &   -0.037 &   76.91328 &   -0.018 \\ 
  57.83881 &   -0.027 &   57.96429 &    0.081 &   58.88613 &   -0.026 &   59.86149 &   -0.020 &   76.91600 &   -0.033 \\ 
  57.83941 &   -0.004 &   57.96688 &    0.106 &   58.90394 &   -0.020 &   59.86358 &   -0.044 &   76.91872 &   -0.008 \\ 
  57.84199 &   -0.015 &   57.96747 &    0.105 &   58.90603 &   -0.038 &   59.93702 &   -0.070 &   76.92144 &   -0.012 \\ 
  57.84259 &    0.000 &   57.96807 &    0.116 &   58.90813 &   -0.034 &   59.93912 &   -0.074 &   76.92416 &   -0.014 \\ 
  57.84319 &   -0.004 &   57.97066 &    0.113 &   58.91023 &   -0.050 &   59.94122 &   -0.055 &   76.92689 &   -0.014 \\ 
  57.84580 &   -0.008 &   57.97126 &    0.127 &   58.91233 &   -0.035 &   59.94537 &   -0.067 &   76.92961 &   -0.006 \\ 
  57.84639 &   -0.012 &   57.97187 &    0.129 &   58.91442 &   -0.031 &   59.94745 &   -0.066 &   76.93233 &   -0.009 \\ 
  57.84699 &   -0.016 &   57.97446 &    0.111 &   58.91652 &   -0.025 &   59.94955 &   -0.068 &   76.93506 &   -0.023 \\ 
  57.84958 &   -0.011 &   57.97506 &    0.076 &   58.91862 &   -0.038 &   59.95163 &   -0.081 &   76.93778 &   -0.013 \\ 
  57.85018 &    0.013 &   57.97565 &    0.064 &   58.92072 &   -0.037 &   59.95371 &   -0.075 &   76.94049 &   -0.022 \\ 
  57.85078 &   -0.026 &   57.97823 &    0.067 &   58.92490 &   -0.036 &   59.95581 &   -0.081 &   76.94318 &   -0.024 \\ 
  57.85338 &   -0.005 &   57.97882 &    0.061 &   58.92699 &   -0.044 &   59.95790 &   -0.081 &   76.94586 &   -0.003 \\ 
  57.85397 &   -0.006 &   57.97943 &    0.058 &   58.92908 &   -0.022 &   59.95999 &   -0.066 &   76.94859 &   -0.013 \\ 
  57.85457 &   -0.024 &   57.98199 &    0.038 &   58.93117 &   -0.027 &   59.96207 &   -0.065 &   76.95131 &   -0.018 \\ 
  57.85716 &   -0.015 &   57.98258 &    0.043 &   58.93327 &   -0.024 &   59.96416 &   -0.063 &   76.95403 &   -0.013 \\ 
  57.85776 &   -0.020 &   57.98318 &    0.026 &   59.64748 &   -0.043 &   59.96625 &   -0.066 &   76.95675 &   -0.025 \\ 
  57.85837 &   -0.028 &   57.98577 &    0.034 &   59.65218 &   -0.016 &   59.96833 &   -0.075 &   76.95947 &   -0.013 \\ 
  57.86096 &   -0.016 &   57.98636 &   -0.005 &   59.65572 &   -0.034 &   59.97043 &   -0.072 &   76.96218 &   -0.014 \\ 
  57.86155 &   -0.040 &   57.98696 &    0.004 &   59.65781 &   -0.027 &   59.97251 &   -0.078 &   76.96490 &   -0.016 \\ 
  57.86215 &   -0.016 &   57.98955 &   -0.025 &   59.65990 &   -0.015 &   59.97461 &   -0.103 &   76.96762 &   -0.020 \\ 
  57.86474 &   -0.023 &   57.99014 &   -0.015 &   59.66198 &   -0.058 &   59.97670 &   -0.076 &   76.97035 &   -0.016 \\ 
  57.86534 &   -0.037 &   57.99074 &   -0.005 &   59.66407 &   -0.034 &   59.97879 &   -0.070 &   76.97307 &   -0.016 \\ 
  57.86594 &   -0.024 &   58.69591 &   -0.032 &   59.66617 &   -0.046 &   59.98088 &   -0.050 &   76.97579 &   -0.026 \\ 
  57.86850 &   -0.002 &   58.69978 &   -0.029 &   59.66826 &   -0.009 &   59.98296 &   -0.057 &   76.97852 &   -0.014 \\ 
\enddata
\end{deluxetable}{}

\begin{deluxetable}{cccccccccc}
\tablecaption{Photometric $I$ Data for LP133-373\label{tbl4}} 
\tablewidth{0pt}
\tablehead{
\colhead{HJD} & \colhead{$\Delta I$} & \colhead{HJD} &  \colhead{$\Delta I$}  & \colhead{HJD} & \colhead{$\Delta I$} & \colhead{HJD} & \colhead{$\Delta I$} & \colhead{HJD} & \colhead{$\Delta I$} \\
 \colhead{-2453800} &          & \colhead{-2453800} &               & \colhead{-2453800} &            & \colhead{-2453800} &                & \colhead{-2453800} &      \\
}
\startdata                        
  57.74927 &   -0.010 &   57.87407 &   -0.018 &   58.70605 &   -0.037 &   59.66716 &   -0.003 &   59.97978 &   -0.060 \\
  57.74988 &   -0.010 &   57.87467 &   -0.009 &   58.70816 &   -0.043 &   59.66925 &   -0.008 &   59.98186 &   -0.068 \\ 
  57.75048 &   -0.030 &   57.87526 &    0.003 &   58.71025 &   -0.049 &   59.67134 &   -0.057 &   59.98396 &   -0.067 \\ 
  57.75305 &   -0.030 &   57.87782 &   -0.001 &   58.71235 &   -0.032 &   59.67341 &   -0.022 &   59.98605 &   -0.058 \\ 
  57.75364 &   -0.011 &   57.87842 &   -0.019 &   58.71443 &   -0.040 &   59.67551 &   -0.002 &   59.98813 &   -0.052 \\ 
  57.75424 &   -0.008 &   57.87901 &   -0.005 &   58.71653 &   -0.049 &   59.67759 &   -0.028 &   59.99022 &   -0.067 \\ 
  57.75683 &   -0.013 &   57.88158 &   -0.013 &   58.71862 &   -0.053 &   59.67969 &   -0.031 &   59.99231 &   -0.047 \\ 
  57.75803 &   -0.015 &   57.88218 &   -0.018 &   58.72072 &   -0.036 &   59.68178 &   -0.012 &   59.99441 &   -0.033 \\ 
  57.76063 &   -0.003 &   57.88279 &   -0.008 &   58.72281 &   -0.040 &   59.68387 &   -0.029 &   76.66179 &   -0.003 \\ 
  57.76122 &   -0.024 &   57.88538 &   -0.019 &   58.72491 &   -0.036 &   59.68595 &   -0.024 &   76.66451 &   -0.004 \\ 
  57.76181 &   -0.009 &   57.88598 &   -0.003 &   58.72700 &   -0.034 &   59.68803 &   -0.025 &   76.66720 &    0.013 \\ 
  57.76438 &   -0.012 &   57.88657 &   -0.001 &   58.72910 &   -0.052 &   59.69013 &   -0.015 &   76.66993 &    0.026 \\ 
  57.76497 &   -0.031 &   57.88917 &   -0.010 &   58.73119 &   -0.031 &   59.69221 &   -0.029 &   76.67264 &    0.032 \\ 
  57.76556 &   -0.022 &   57.88978 &    0.001 &   58.73329 &   -0.031 &   59.69429 &   -0.029 &   76.67537 &    0.065 \\ 
  57.76813 &    0.007 &   57.89037 &    0.002 &   58.73537 &   -0.035 &   59.69638 &   -0.008 &   76.67808 &    0.082 \\ 
  57.76872 &   -0.005 &   57.89296 &   -0.008 &   58.73746 &   -0.040 &   59.69847 &   -0.024 &   76.68082 &    0.095 \\ 
  57.76931 &   -0.003 &   57.89356 &   -0.016 &   58.73956 &   -0.032 &   59.70055 &   -0.033 &   76.68352 &    0.105 \\ 
  57.77249 &   -0.002 &   57.89415 &   -0.012 &   58.74165 &   -0.030 &   59.70265 &   -0.039 &   76.68624 &    0.118 \\ 
  57.77308 &   -0.014 &   57.89676 &   -0.010 &   58.74375 &   -0.026 &   59.70474 &   -0.033 &   76.68897 &    0.116 \\ 
  57.77567 &   -0.015 &   57.89734 &   -0.023 &   58.74583 &   -0.028 &   59.70684 &   -0.050 &   76.69170 &    0.119 \\ 
  57.77627 &    0.000 &   57.89795 &   -0.012 &   58.74793 &   -0.038 &   59.70892 &   -0.029 &   76.69443 &    0.116 \\ 
  57.77688 &    0.003 &   57.90054 &   -0.001 &   58.75002 &   -0.049 &   59.71102 &   -0.028 &   76.69715 &    0.096 \\ 
  57.77947 &   -0.015 &   57.90114 &    0.002 &   58.75212 &   -0.038 &   59.71311 &   -0.019 &   76.69987 &    0.079 \\ 
  57.78006 &   -0.024 &   57.90173 &   -0.020 &   58.75421 &   -0.039 &   59.71519 &   -0.031 &   76.70259 &    0.061 \\ 
  57.78066 &   -0.028 &   57.90434 &   -0.010 &   58.75632 &   -0.040 &   59.71728 &    0.014 &   76.70530 &    0.036  \\
  57.78325 &   -0.024 &   57.90494 &   -0.016 &   58.75841 &   -0.051 &   59.71935 &   -0.040 &   76.70803 &    0.020 \\ 
  57.78384 &   -0.017 &   57.90553 &   -0.004 &   58.76051 &   -0.017 &   59.72144 &   -0.027 &   76.71073 &    0.008 \\ 
  57.78444 &   -0.034 &   57.90812 &   -0.001 &   58.76260 &   -0.014 &   59.72354 &   -0.016 &   76.71346 &   -0.009 \\ 
  57.78703 &   -0.006 &   57.90872 &   -0.014 &   58.76468 &   -0.011 &   59.72562 &   -0.035 &   76.71618 &   -0.004 \\ 
  57.78762 &   -0.017 &   57.90932 &   -0.002 &   58.76678 &    0.012 &   59.72772 &   -0.025 &   76.71890 &   -0.006 \\ 
  57.78822 &   -0.018 &   57.91193 &   -0.002 &   58.76889 &    0.041 &   59.72980 &   -0.019 &   76.72163 &   -0.015 \\ 
  57.79079 &   -0.030 &   57.91252 &    0.012 &   58.77098 &    0.044 &   59.73190 &   -0.026 &   76.72795 &   -0.001 \\ 
  57.79139 &   -0.028 &   57.91312 &   -0.011 &   58.77307 &    0.068 &   59.73399 &   -0.035 &   76.73401 &    0.002 \\ 
  57.79198 &   -0.003 &   57.91571 &   -0.013 &   58.80441 &   -0.042 &   59.73607 &   -0.018 &   76.76668 &   -0.015 \\ 
  57.79457 &   -0.019 &   57.91630 &   -0.016 &   58.80650 &   -0.035 &   59.73817 &   -0.026 &   76.76939 &   -0.005 \\ 
  57.79517 &   -0.002 &   57.91690 &   -0.008 &   58.80859 &   -0.020 &   59.74026 &   -0.022 &   76.77208 &   -0.016 \\ 
  57.79576 &   -0.016 &   57.91950 &   -0.010 &   58.81069 &   -0.039 &   59.77498 &   -0.062 &   76.77478 &   -0.024 \\ 
  57.79837 &   -0.020 &   57.92010 &   -0.018 &   58.81278 &   -0.033 &   59.78040 &   -0.049 &   76.77750 &   -0.010 \\ 
  57.79896 &   -0.030 &   57.92070 &   -0.016 &   58.81488 &   -0.031 &   59.78318 &   -0.031 &   76.78022 &   -0.013 \\ 
  57.79956 &   -0.006 &   57.92329 &   -0.005 &   58.81697 &   -0.038 &   59.78526 &   -0.034 &   76.78295 &   -0.028 \\ 
  57.80216 &   -0.017 &   57.92388 &   -0.014 &   58.81908 &   -0.035 &   59.78734 &   -0.046 &   76.78566 &   -0.027 \\ 
  57.80277 &   -0.017 &   57.92448 &   -0.013 &   58.82118 &   -0.028 &   59.78944 &   -0.031 &   76.78838 &   -0.017 \\ 
  57.80336 &   -0.018 &   57.92708 &   -0.006 &   58.82327 &   -0.031 &   59.79153 &   -0.032 &   76.79109 &   -0.002 \\ 
  57.80595 &   -0.024 &   57.92768 &   -0.005 &   58.82536 &   -0.026 &   59.79362 &   -0.038 &   76.79382 &   -0.019 \\ 
  57.80655 &   -0.004 &   57.92828 &    0.001 &   58.82746 &   -0.024 &   59.79571 &   -0.037 &   76.79653 &   -0.010 \\ 
  57.80715 &   -0.024 &   57.93086 &   -0.012 &   58.82954 &   -0.037 &   59.79779 &   -0.042 &   76.79925 &   -0.015 \\ 
  57.80975 &   -0.022 &   57.93145 &    0.004 &   58.83164 &   -0.034 &   59.79989 &   -0.035 &   76.80199 &   -0.017 \\ 
  57.81035 &   -0.031 &   57.93204 &    0.008 &   58.83374 &   -0.033 &   59.80198 &   -0.044 &   76.80471 &   -0.015 \\ 
  57.81095 &   -0.014 &   57.93462 &   -0.008 &   58.83581 &   -0.035 &   59.80407 &   -0.019 &   76.80743 &   -0.018 \\ 
  57.81354 &   -0.012 &   57.93523 &   -0.018 &   58.83791 &   -0.038 &   59.80614 &   -0.013 &   76.81015 &   -0.024 \\ 
  57.81414 &   -0.004 &   57.93583 &    0.009 &   58.84001 &   -0.040 &   59.80821 &   -0.019 &   76.81287 &   -0.022 \\ 
  57.81475 &    0.003 &   57.93842 &   -0.005 &   58.84210 &   -0.030 &   59.81028 &   -0.029 &   76.81558 &   -0.025 \\ 
  57.81734 &   -0.020 &   57.93901 &   -0.016 &   58.84421 &   -0.037 &   59.81238 &   -0.033 &   76.81830 &   -0.027 \\ 
  57.81794 &   -0.016 &   57.93961 &   -0.017 &   58.84630 &   -0.031 &   59.81447 &   -0.028 &   76.82102 &   -0.020 \\ 
  57.81854 &   -0.011 &   57.94222 &   -0.011 &   58.84840 &   -0.039 &   59.81655 &   -0.033 &   76.82374 &   -0.011 \\ 
  57.82113 &   -0.018 &   57.94281 &   -0.018 &   58.85049 &   -0.026 &   59.81865 &   -0.039 &   76.82646 &   -0.018 \\ 
  57.82173 &   -0.029 &   57.94341 &   -0.012 &   58.85258 &   -0.026 &   59.82073 &   -0.036 &   76.82919 &   -0.017 \\ 
  57.82233 &   -0.009 &   57.94599 &    0.004 &   58.85469 &   -0.015 &   59.82283 &   -0.027 &   76.83191 &   -0.013 \\ 
  57.82491 &   -0.019 &   57.94659 &    0.010 &   58.85679 &   -0.031 &   59.82491 &   -0.034 &   76.87372 &   -0.018 \\ 
  57.82551 &   -0.019 &   57.94719 &    0.009 &   58.85888 &   -0.034 &   59.82701 &   -0.055 &   76.87644 &   -0.023 \\ 
  57.82610 &   -0.005 &   57.94977 &    0.027 &   58.86098 &   -0.034 &   59.82910 &   -0.036 &   76.87916 &   -0.006 \\ 
  57.82869 &   -0.005 &   57.95038 &    0.022 &   58.86307 &   -0.023 &   59.83118 &   -0.037 &   76.88460 &   -0.053 \\ 
  57.82929 &   -0.016 &   57.95098 &    0.026 &   58.86516 &   -0.037 &   59.83327 &   -0.033 &   76.88733 &   -0.031 \\ 
  57.82988 &   -0.014 &   57.95357 &    0.033 &   58.86725 &   -0.031 &   59.83535 &   -0.024 &   76.89006 &   -0.027 \\ 
  57.83248 &   -0.015 &   57.95417 &    0.045 &   58.86934 &   -0.020 &   59.83745 &   -0.017 &   76.89279 &   -0.026 \\ 
  57.83308 &   -0.009 &   57.95476 &    0.050 &   58.87145 &   -0.031 &   59.83953 &   -0.047 &   76.89552 &   -0.019 \\ 
  57.83367 &   -0.013 &   57.95735 &    0.086 &   58.87354 &   -0.039 &   59.84161 &   -0.020 &   76.89824 &   -0.031 \\ 
  57.83626 &   -0.016 &   57.95794 &    0.081 &   58.87565 &   -0.028 &   59.84369 &   -0.028 &   76.90097 &   -0.024 \\ 
  57.83686 &   -0.019 &   57.95855 &    0.094 &   58.87775 &   -0.032 &   59.84579 &   -0.036 &   76.90369 &   -0.028 \\ 
  57.83746 &   -0.013 &   57.96114 &    0.095 &   58.87984 &   -0.023 &   59.84788 &   -0.017 &   76.90641 &   -0.005 \\ 
  57.84004 &   -0.013 &   57.96174 &    0.078 &   58.88192 &   -0.025 &   59.84997 &   -0.040 &   76.90914 &   -0.011 \\ 
  57.84065 &   -0.018 &   57.96234 &    0.089 &   58.88403 &   -0.026 &   59.85205 &   -0.032 &   76.91186 &   -0.018 \\ 
  57.84125 &   -0.031 &   57.96492 &    0.114 &   58.88613 &   -0.026 &   59.85415 &   -0.023 &   76.91459 &   -0.014 \\ 
  57.84384 &   -0.023 &   57.96552 &    0.094 &   58.90394 &   -0.020 &   59.85623 &   -0.036 &   76.91731 &   -0.022 \\ 
  57.84444 &   -0.017 &   57.96613 &    0.100 &   58.90603 &   -0.038 &   59.85831 &   -0.022 &   76.92003 &   -0.010 \\ 
  57.84504 &   -0.012 &   57.96872 &    0.116 &   58.90813 &   -0.034 &   59.86039 &   -0.050 &   76.92275 &   -0.014 \\ 
  57.84764 &   -0.024 &   57.96932 &    0.120 &   58.91023 &   -0.050 &   59.86248 &   -0.032 &   76.92547 &   -0.018 \\ 
  57.84823 &    0.003 &   57.96992 &    0.094 &   58.91233 &   -0.035 &   59.93802 &   -0.060 &   76.92819 &   -0.010 \\ 
  57.84883 &   -0.005 &   57.97250 &    0.113 &   58.91442 &   -0.031 &   59.94012 &   -0.069 &   76.93092 &   -0.024 \\ 
  57.85143 &   -0.018 &   57.97310 &    0.100 &   58.91652 &   -0.025 &   59.94221 &   -0.055 &   76.93364 &   -0.011 \\ 
  57.85202 &   -0.022 &   57.97371 &    0.106 &   58.91862 &   -0.038 &   59.94428 &   -0.059 &   76.93636 &   -0.014 \\ 
  57.85262 &   -0.029 &   57.97630 &    0.035 &   58.92072 &   -0.037 &   59.94635 &   -0.060 &   76.93909 &   -0.013 \\ 
  57.85522 &   -0.032 &   57.97690 &    0.091 &   58.92490 &   -0.036 &   59.94845 &   -0.066 &   76.94179 &   -0.012 \\ 
  57.85582 &   -0.013 &   57.97749 &    0.053 &   58.92699 &   -0.044 &   59.95053 &   -0.069 &   76.94447 &   -0.011 \\ 
  57.85642 &   -0.017 &   57.98006 &    0.094 &   58.92908 &   -0.022 &   59.95263 &   -0.060 &   76.94717 &   -0.011 \\ 
  57.85900 &   -0.013 &   57.98065 &    0.072 &   58.93117 &   -0.027 &   59.95471 &   -0.075 &   76.94989 &   -0.005 \\ 
  57.85960 &    0.000 &   57.98125 &    0.050 &   58.93327 &   -0.024 &   59.95680 &   -0.059 &   76.95262 &   -0.015 \\ 
  57.86021 &   -0.020 &   57.98381 &    0.030 &   58.93536 &   -0.012 &   59.95889 &   -0.057 &   76.95534 &   -0.018 \\ 
  57.86280 &   -0.024 &   57.98441 &    0.021 &   58.93745 &   -0.060 &   59.96098 &   -0.059 &   76.95806 &   -0.027 \\ 
  57.86340 &   -0.012 &   57.98502 &    0.032 &   58.93955 &   -0.054 &   59.96307 &   -0.065 &   76.96078 &   -0.028 \\ 
  57.86399 &   -0.019 &   57.98761 &    0.027 &   58.94762 &   -0.027 &   59.96515 &   -0.064 &   76.96348 &   -0.012 \\ 
  57.86657 &   -0.031 &   57.98821 &    0.016 &   58.94970 &   -0.016 &   59.96724 &   -0.078 &   76.96621 &   -0.003 \\ 
  57.86717 &   -0.014 &   57.98880 &    0.000 &   58.95181 &   -0.072 &   59.96933 &   -0.075 &   76.96893 &   -0.013 \\ 
  57.86776 &   -0.009 &   58.69591 &   -0.032 &   59.65880 &   -0.037 &   59.97142 &   -0.079 &   76.97166 &   -0.016 \\ 
  57.87032 &   -0.022 &   58.69978 &   -0.029 &   59.66089 &    0.000 &   59.97351 &   -0.060 &   76.97438 &   -0.019 \\ 
  57.87091 &   -0.017 &   58.70187 &   -0.027 &   59.66297 &   -0.064 &   59.97560 &   -0.056 &   76.97710 &   -0.027 \\ 
  57.87151 &   -0.015 &   58.70397 &   -0.042 &   59.66507 &   -0.042 &   59.97768 &   -0.056 &   76.97983 &   -0.017 \\ 
\enddata
\end{deluxetable}{}                        

\clearpage

\end{document}